\documentclass[12pt]{article}

\usepackage[margin=.8in]{geometry}

\usepackage{setspace}

\usepackage{amssymb}
\usepackage{amsmath}
\usepackage{graphicx}
\usepackage{mathdots}
\usepackage{slashed}
\usepackage{verbatim}
\usepackage{xcolor}
\usepackage{cancel}

\usepackage{cite}

\usepackage{hyperref}

\usepackage{IEEEtrantools}

\usepackage{amsfonts}

\usepackage{epsfig}

\newcommand{\be}{\begin{equation}}
\newcommand{\ee}{\end{equation}}
\newcommand{\bea}{\begin{eqnarray}}
\newcommand{\eea}{\end{eqnarray}}
\newcommand{\ba}{\begin{eqnarray}}
\newcommand{\ea}{\end{eqnarray}}

\def\a{\alpha }
\def\EE{E}
\def\s{\sigma }
\def\p{\partial}
\def\non{\nonumber}

\begin{document}


\mbox{}
\vspace{0truecm}
\linespread{1.1}

\vspace{0.5truecm}

\centerline{\large \bf  Probing Hidden Dimensions via Muon Lifetime Measurements\footnote{This is an expanded and detailed version of an essay that received an honorable Mention in the Gravity Research Foundation 2024 Awards.}} 



\vspace{1truecm}

\centerline{
    {\bf Jorge G. Russo} }

\vspace{0.5cm}

\noindent  
\centerline {\it Instituci\'o Catalana de Recerca i Estudis Avan\c{c}ats (ICREA), }
\centerline{\it Pg. Lluis Companys, 23, 08010 Barcelona, Spain.}

\medskip
\noindent 
\centerline{\it  Departament de F\' \i sica Qu\' antica i Astrof\'\i sica and Institut de Ci\`encies del Cosmos,}
\centerline{\it Universitat de Barcelona, Mart\'i Franqu\`es, 1, 08028 Barcelona, Spain. }

\medskip

\centerline{  {\it E-Mail:}  {\texttt jorge.russo@icrea.cat} }

\vspace{1cm}

\centerline{\bf ABSTRACT}
\medskip

In the context of Kaluza-Klein theories, the time dilation of charged particles in an external field
depends on the charge in a specific way. 
Experimental tests are proposed to search for extra dimensions using this distinctive feature.

\bigskip\medskip

 \bigskip
 \bigskip
 
\centerline{October, 2024.}

\newpage








\setcounter{equation}{0}

Since the remarkable hypothesis of Theodor Kaluza in 1921 \cite{Kaluza:1921tu},
further developed by Klein \cite{Klein:1926tv},
extra dimensions have played a fundamental role in many
unified theories such as superstring theory.
Although several experiments have been proposed to detect Kaluza-Klein (KK) resonances, there is currently no observational evidence of extra dimensions.
This paper explores a novel strategy for searching for extra dimensions based on the lifetime of charged  particles in the presence of a KK external field.


To simplify the setting,  we will first consider a five-dimensional spacetime with the topology
$\mathbb{R}^4\times \mathbb{S}^1$, where $\mathbb{S}^1$ is parameterized by a periodic coordinate $y\approx y+2\pi R$.
The five-dimensional theory is assumed to be  Einstein theory coupled to a massive scalar field. The dimensional reduction to four dimensions will give rise to gravity coupled to
a KK vector field $A_\mu$ and a dilaton $\phi$, coupled to a massive scalar field and to a tower of
KK particles.

In this study, we will assume that the compactification radius $R$ is much smaller than all relevant scales in the problem. This is required by the dimensional reduction ansatz and it ensures that the low-energy effective theory is described by $A_\mu$, $\phi$ and the four-dimensional metric.

Our starting point is thus a purely gravitational system where there are no gauge fields (and therefore no charged
particles). 
Consider a clock in an arbitrary gravitational field in the five-dimensional theory. The time dilation is simply given by the formula \cite{Weinberg:1972kfs}
\be
\frac{d \tau}{dt} =\left( -g_{AB} \frac{dx^A}{dt} \frac{dx^B}{dt}\right)^{1/2}\ ,\qquad A,B=0,...,4\ .
\label{uno}
\ee
Here $d \tau$ represents the period between ticks when the clock is at rest in the absence of gravitation,
and $dt$ is the time interval between ticks when the clock is placed on a gravitational field $g_{\mu\nu}$ and moves with velocity $dx^A/dt$. 

Let us now investigate this formula for a specific gravitational field
that leads to a non-vanishing vector field in four dimensions.
We consider the $d=4+1$ metric that upon reduction leads to an electrically charged black hole. The solution can be obtained by boosting the Schwarzschild solution in the $y$ direction \cite{Gibbons:1985ac}.
It is given by 
\bea
ds_5^2 &=& - \frac{1}{ 1+\frac{\a}{ r}} \Big( 1-\frac{r_h}{ r} - \frac{Q^2}{ r^2}\Big) dt^2 +\frac{dr^2}{ 1-\frac{r_h}{ r}}
\non\\
&+& \big( 1+\frac{\a}{ r} \big) dy^2 -2\frac{Q}{r} dy dt 
+r^2 d\Omega_2^2 \ .
\label{dos}
\eea
It is a solution of the $d=4+1$ vacuum Einstein equation $R_{\mu\nu} =0$
provided
\be
Q^2=\a(\a+r_h)\ .
\label{tres}
\ee
Here we choose $\a >0$.
The dimensional reduction is carried out by the ansatz
\be
ds^2_5=e^{\frac{2\phi}{\sqrt{3}}}\hat g_{\mu\nu}+e^{-\frac{4\phi}{\sqrt{3}}}(dy-A_\mu dx^\mu)^2 \ .
\ee
with 
$$
\hat g_{\mu\nu} = e^{-{2 \phi\over\sqrt{3} } }\left(g_{\mu\nu}- g_{\mu y} g_{\nu y}\right), \ \ \  A_\mu \equiv {g_{\mu y} \over g_{yy}}\ ,\quad g_{yy}=e^{-\frac{4\phi}{\sqrt{3}}}\ .
$$
Here we have set $\kappa^2\equiv 4\pi G =1$. The electromagnetic field with canonical normalization
is obtained by $A_\mu\to 2\kappa A_\mu$ (for a discussion, see  \cite{Gibbons:1985ac}).

The dimensional reduction gives the four-dimensional metric $\hat g_{\mu\nu}$ in the Einstein frame, along with a
KK electric field and a KK scalar field $\phi $ 
\bea
ds^2_4 &=&  - \frac{1-\frac{r_h}{r} }{\big(1+\frac{\a}{ r}\big)^{\frac{1}{2}}}  dt^2 +\frac{dr^2}{ 1-\frac{r_h}{ r}}
\big(1+\frac{\a}{ r}\big)^{\frac{1}{2}} 
+r^2 \big(1+\frac{\a}{ r}\big)^{\frac{1}{2}}  d\Omega_2^2 \ ,
\label{cuatro}\\
F_{rt}^2 &=& -\frac{Q^2}{  \ (r+\a)^4}\ ,\qquad \phi =-\frac{\sqrt{3}}{ 4} \log \big( 1+\frac{\a}{ r}\big)\ .
\label{cinco}
\eea
It should be noted that this geometry  not only describes black holes but also the gravitational field
for any central source with spherical symmetry and charge $Q$ \cite{Gibbons:1985ac}.

The event horizon is at $r=r_h$ and there is a curvature singularity
at $r=0$. A detailed discussion on the properties of this solution can be found in 
\cite{Gibbons:1985ac} (the radial coordinate here corresponds to $r-r_-$
in the notation of \cite{Gibbons:1985ac}).
The ADM mass $M$ is
\be
2MG= r_h+\frac{\a}{ 2}\ .
\ee
Using these relations one can express $r_h$ and $\alpha $ in terms of $Q$ and $M$,
\be
r_h=3MG-\sqrt{M^2G^2+\frac{Q^2}{2}}\ ,\qquad \alpha =\frac12\left(\sqrt{M^2G^2+\frac{Q^2}{2}}-MG\right)\ .
\ee
Demanding $r_h> 0$ leads to the condition $Q< 4MG$. 
For $Q=4MG$, one has $r_h=0$, $\alpha =4MG$ and the four-dimensional metric becomes singular.

Now consider a massive  particle moving along a radial geodesic
in the five-dimensional geometry (\ref{dos}).
According to (\ref{uno}), it should undergo a
time dilation  given by
\be
\frac{d\tau }{dt} =\left(  \frac{1-\frac{r_h}{r} }{ 1+\frac{\a}{r}}  -\frac{\dot r^2}{ 1-\frac{r_h}{r}}
- \big( 1+\frac{\a}{r} \big) \left(\dot y -\frac{Q}{ r+\a } \right)^2  \right)^{1/2}\ .
\label{siete}
\ee
In the four-dimensional picture, 
a naive application of (\ref{uno}) using the geometry  (\ref{cuatro}) would  give, for the same particle, a different factor,
\be
\frac{d \tau}{ dt} =\left( \frac{1-\frac{r_h}{r}}{ \big(1+\frac{\a}{r}\big)^{\frac12 }}  - 
\frac{\dot r^2}{ 1-\frac{r_h}{r}}\big(1+\frac{\a}{r}\big)^{\frac12}  \right)^{\frac12}\ .
\label{seis}
\ee
This is not the correct time dilation factor, and the reason is due to the different nature of the probes with five-dimensional origin.
The probe here is a KK particle and it carries the  couplings to $\phi$ and $A_\mu$ inherited from the dimensional reduction.

The time dilation factor is uniquely determined from the gravitational interactions in five dimensions. 
For a particle of mass $m$ in five dimensions, the action  is 
\be
S= -m\int d\lambda \ \sqrt{- g_{AB} \dot x^A \dot x^B } \ .
\label{ggg}
\ee
Let us consider the ansatz
$t=\lambda ,\ r=r(t) ,\ y=y(t)$, describing radially moving particles with
momentum in the internal direction. Then
\be
S= -m\int dt\ \Delta \ ,
\label{accion}
\ee
\be
\Delta \equiv \sqrt{ \frac{1-\frac{r_h}{r} }{ 1+\frac{\a}{r}}  -\frac{\dot r^2}{ 1-\frac{r_h}{r}}
- \big( 1+\frac{\a}{r} \big) \big(\dot y - \frac{Q}{ r+\a }  \big)^2
}\ .
\nonumber
\ee
Since the metric does not depend on $y$, its conjugate momentum $P_y$ is conserved:
\be
P_y =  \frac{m }{ \Delta } \big( 1+\frac{\a}{r} \big) (\dot y -\frac{Q}{ r+\a } \big)={\rm const.}
\label{pey}
\ee
Because of the periodic nature of the coordinate $y$, one has the usual quantization $P_y=n/R$, where $n$ is an integer.
At the same time, also energy is conserved. This gives
\be
\EE  \equiv \frac{m}{\Delta } \left( h(r) + \frac{Q}{r}\dot  y \right)\ ,\qquad  h(r)\equiv \frac{1-\frac{r_h}{r} -\frac{Q^2}{r^2}}{ 1+\frac{\a}{r}}\ .
\label{energia}
\ee
Hence
\be
\dot y = \frac{P_y}{ \EE }\ \frac{h(r)+\frac{\EE Q}{ P_y r}}{ 1+\frac{\a}{r} -\frac{QP_y}{ \EE  r} }\ .
\label{solu}
\ee
Substituting this into eq. (\ref{energia}), one finds a long expression that can be solved for $\dot r $. 
Integrating this, one can find $r=r(t)$. Here we only need $\Delta $, which has a simple expression:
\be
\Delta = \frac{m}{ \EE  }\  \frac{1-\frac{r_h}{r}} {1+(\a-\frac{QP_y}{ \EE })\frac{1}{ r} }\ .
\ee
The dynamics is  characterized by the two parameters $P_y $ and $\EE $.
There are two possible situations: either the particle gets to $r=\infty $, or it gets up to some $r_{\rm max} $ where $\dot r =0$.
The condition to get to infinity is
\be
\dot r_\infty^2>0\ \to \ \EE  > \sqrt{m^2+P_y^2}\ .
\label{condi}
\ee
In this case, $\dot y_\infty =P_y/\EE $ and
\be
\EE  = \frac{m}{ \sqrt{1- \vec v^2} }\ ,\quad \vec v^2= \dot r_\infty^2+ \dot y_\infty ^2\ .
\ee
Using  (\ref{uno}), the time dilation factor for a particle following this geodesic is thus given by
\be
\frac{d \tau}{dt} = \Delta = \frac{m}{\EE}\  \frac{1-\frac{r_h}{r}}{1+(\a-\frac{QP_y}{\EE})\frac{1}{r} }\ ,\qquad \ r=r(t)\ .
\label{guar}
\ee

\medskip

It is instructive to  reproduce  this formula in the 3+1 dimensional context.
First of all, we need to find the dimensional reduction of the particle action (\ref{ggg}).
Introducing a Lagrange multiplier $\ell $, the action (\ref{ggg}) is equivalent to
\bea
S &=&  \frac12 \int d\lambda  \Big( \frac{1}{ \ell} g_{AB} \dot x^A\dot x^B - m^2 \ell \Big)
\non\\
&=&  \frac12 \int d\lambda \Big( \frac{1}{ \ell} G_{\mu\nu} \dot x^\mu\dot x^\nu + 
\frac{1}{ \ell} g_{yy}\big(\dot y + A_\mu \dot x^\mu )^2- m^2 \ell \Big) \ ,
\non
\eea
where 
$$
G_{\mu\nu} \equiv g_{\mu\nu}- \frac{g_{\mu y} g_{\nu y} }{ g_{yy}}\ ,\qquad A_\mu \equiv \frac{g_{\mu y} }{ g_{yy}}
\ ,\qquad \mu=0,...,3\ .
$$
Now introduce $P_y$ as a Lagrange multiplier:
\be
S =  \frac12\int d\lambda \Big( \frac{1}{ \ell} G_{\mu\nu} \dot x^\mu\dot x^\nu -\frac{\ell P_y^2}{  g_{yy}} +
2 P_y \big(\dot y + A_\mu \dot x^\mu )- m^2 \ell \Big) \ .
\ee
The $y$ equation of motion gives the conservation law $\dot P_y=0$, where we assume that
the metric does not depend on $y$ as  required by the Kaluza-Klein reduction.
Then the term $P_y\dot y$ becomes a total derivative and does not
contribute to the equations of motion.

Eliminating  $\ell $ through its equation of motion, we find
\be
S = - \int d\lambda \Big(   m_4(x) e^{\frac{\phi}{\sqrt{3}}} \ \sqrt{- \hat g_{\mu\nu} \dot x^\mu\dot x^\nu} 
-(\dot y+ P_y A_\mu \dot x^\mu) \Big)\ ,
\label{ccc}
\ee
where  $\hat g_{\mu\nu} = e^{-\frac{2 \phi}{\sqrt{3} } } G_{\mu\nu} $ and
\be
m_4^2 \equiv m^2 +\frac{P_y^2}{g_{yy}} \ .
\ee
represents the four-dimensional mass.

The field-theory counterpart  can be deduced
by dimensional reduction of the field theory action of a scalar field $\s $ of mass $m$.
The starting point is 
\be
S_{\rm FT}= \int d^5 x \sqrt{- g} \big( g^{AB} \p_A\sigma   \p_B \sigma  + m^2\sigma^2 \big)\ .
\ee
Assuming $\sigma(x^\mu,y)=\rho(x^\mu ) e^{i P_y y}$, then dimensional reduction leads to 
\be
S_{\rm FT} = \int d^4 x \sqrt{-\hat g} \big( \hat g^{\mu\nu} D_\mu^*\rho   D_\nu \rho  + m_4^2(x) e^{\frac{2\phi}{\sqrt{3}}}
\rho^2 \big)\ ,
\label{ftch}
\ee
where $D_\mu=\p_\mu -i P_y A_\mu $, representing a scalar field of effective mass $m_4(x) e^{\frac{\phi}{\sqrt{3}}} $, charge $P_y$,
coupled to  $\hat g_{\mu\nu},\ \phi, \ A_\mu $, in  agreement with the particle action (\ref{ccc}).

The proper time for a KK particle following a  geodesic is given by
\be
\frac{d \tau}{dt} = - \frac{\cal L}{m}
\label{ene}
\ee
evaluated on that geodesic.
In the absence of fields, this becomes
$d \tau= dt\ \sqrt{1-\vec v^2}$.
Now consider the Lagrangian (\ref{ccc}) and a geodesic $t=\lambda,\ r=r(t)$.
Since the system is dynamically equivalent to the 5-dimensional system, the solution for $\dot r$ is as 
before, as can be easily checked. Define $\Delta_4\equiv \sqrt{- G_{\mu\nu} \dot x^\mu\dot x^\nu}$.
Then
\bea
\Delta_4^2 &=&\Delta^2 + \big( 1+\frac{\a}{r} \big) \big(\dot y - \frac{Q}{ r+\a }  \big)^2
\non\\
&=& \Delta^2 \Big(1+ \frac{P_y^2}{ m^2g_{yy}} \Big)
\label{loi}
\eea
where we used (\ref{accion}) and (\ref{pey}).
Thus
\be
\frac{d \tau}{ dt} =   \sqrt{1+ \frac{P_y^2}{m^2g_{yy}} }  \Delta_4 - P_y\big( \dot y + A_\mu \dot x^\mu \big)
= \Delta \ .
\label{enez}
\ee
where (\ref{pey}) and (\ref{loi}) have been used. As expected, we reproduce the time dilation (\ref{guar}) found in five dimensions.

  \medskip

 Let us now investigate the significance of the time dilation factor  (\ref{guar}). 
  Expanding at large $r$ (assuming the condition (\ref{condi}) so that the geodesics gets to $r=\infty $), we find
\bea
\frac{d \tau}{dt} &=& 
\frac{m}{\EE }\left( 1 -  \frac{1}{r} \Big( {r_h}+\a  - \frac{P_y Q}{\EE } \Big)\right)+
O\Big(\frac{1}{r^2}\Big) 
\non\\
&=& 
 \sqrt{1-\vec v^2} \Big(1 - \frac{2MG}{r}   +\frac{P_y}{\EE}\frac{Q}{r} - \frac{\a}{2r}\Big) +
O\Big(\frac{1}{r^2}\Big) \ .
\label{fed}
\eea
The term proportional to $2MG$ is the contribution that one would get in pure Einstein theory without gauge or dilaton fields.
Note that for a static particle (\ref{uno}) gives  a contribution $-MG/r$ instead of $-2MG/r$. The numerical factor in front of $MG/r$ differs because the particle in this case follows a geodesic and $\dot r\neq 0$. The term proportional to the charge $P_y$ is
\be
\left(\frac{P_y}{E}\right) \left(\frac{Q}{r}\right)\sqrt{1-\vec v^2} \ .
\ee
In the dimensionally reduced theory, $P_y/E$ is nothing but the ratio of the charge  to the
energy and $Q/r$ is the electric potential. In five dimensions, this term
has only a gravitational origin and it just follows from
the standard formula (\ref{uno}).
Finally, there is a term $-\alpha/(2r)$, which 
originates from the expansion of the $g_{00}$ component
of the five-dimensional metric.


The time dilation factor  has distinctive footprints
of its Kaluza-Klein origin. In particular, this
factor will dictate how the lifetime of charged KK particles is modified in a gravitational field and in
the presence of an external KK electric field (which is also a gravitational field).
The time rates of a charged particle can be compared at two different values of the radius,
$r=r_0+L$ and $r=r_0$, as the particle falls down following a radial geodesic.
Assume for simplicity,  $\dot r_\infty =0$, in which case $E=\sqrt{m^2+P_y^2}$ represents the four-dimensional mass.
Using (\ref{guar}) we obtain
\be
\frac{\delta t_2}{\delta t_1} =   \frac{1-\frac{r_h}{r_0+L}}{1-\frac{r_h}{r_0} }\ 
 \frac{1+(\a-\frac{QP_y}{\EE})\frac{1}{r_0}}{1+(\a-\frac{QP_y}{\EE})\frac{1}{r_0+L} }\ .
\label{guar2}
\ee
We recall that the spacetime \eqref{cuatro} not only describes the gravitational and electromagnetic fields in a spherically symmetric KK  black hole but also the fields
outside any electrically charged central source with spherical symmetry.
To illustrate, if the source is the Earth, with a given net charge $Q$, then $r_0$ may represent the Earth radius and $r_0+L$ may be interpreted as a height within the atmosphere.

For $r_0\gg r_h$ and $r_0\gg L$, the formula \eqref{guar2} becomes
\be
\frac{\delta t_2}{\delta t_1} \approx  1+\frac{L}{r_0^2}\left( 2MG-\frac{P_y Q}{E}+\frac{\alpha}{2}\right)
\ .
\label{guare}
\ee
%
Let us now examine the implications of the formula (\ref{guare}).
The term $2MGL/r_0^2$ is the standard correction of  four-dimensional Einstein theory, whereas the term $L (P_y/E) (Q /r_0^2)$
is a contribution from the KK electric field $(Q /r_0^2)$.
To make an estimate of these corrections, let us write $Q=4 aMG$, where $0\leq a<1$. The strongest effect will occur for $a\approx 1$
and assuming $P_y<0$ (if the particle charge is positive, one may  choose $Q=-4aMG$).
In this case  $\alpha \approx 4MG$ and
\be
\frac{\delta t_2}{\delta t_1}\bigg|_{Q\approx 4MG} \approx  1+\frac{L}{r_0^2}\left( 4MG+\frac{4|n| MG}{Rm_4}\right)
\ ,
\label{guaree}
\ee
where we have set $P_y=n/R$. We recall that the four dimensional mass is $m_4^2=m^2+n^2/R^2$.
If $mR\ll n$, then the second term approaches 1 and one gets
\be
\frac{\delta t_2}{\delta t_1}\bigg|_{Q\approx 4MG} \approx  1+\frac{L}{r_0^2} 8MG\ ,\qquad mR\ll n\ .
\label{guarees}
\ee
Thus we see that the standard gravitational effect of pure four-dimensional Einstein theory has been multiplied
by a factor of 4. 
More generally, as the charge $Q$ is gradually increased
from 0 to $4MG$, the relative factor  increases from 0 to 4.
It is important to recall that, from the five-dimensional perspective, the time dilation \eqref{guarees} is only
 due to   gravitational effects, since the five-dimensional theory is pure Einstein theory.
The above time rate is strictly dictated  by the gravitational field.

If the electromagnetic field has a KK origin, one
can anticipate that the correction to the time rate in \eqref{guare} originating from an electric field $Q/r_0^2$
will  occur not only in black holes but also in any electric field applied in an Earth-based laboratory. 
 The reason is that in Kaluza-Klein theories $A_\mu$ is a component of the metric, $A_\mu=g_{\mu y}/g_{yy}$, and this component enters into the time dilation formula \eqref{uno}, therefore affecting the time rate.
 Consider a particle with Kaluza-Klein charge $n$ moving in a uniform electric field ${\bf E}$, with vector potential $A_0={\bf E}\cdot {\bf x}$.
 From the five-dimensional standpoint, this particle follows a geodesics.
 Ignoring gravitational (and dilaton) back reaction, the action is 
 \be
S= -m\int dt\ \Delta \ ,
\nonumber
\ee
\be
\Delta \equiv \sqrt{ 1- \dot {\bf x}^2 -\big(\dot y - A_0  \big)^2
}\ .
\label{acciones}
\ee
As usual, the back-reaction can be ignored, provided that 
$8\pi G |{\bf E}|^2\ll 1$, an approximation that holds with great accuracy for all electric fields produced in standard laboratory experiments.

 Using that the momentum $P_y$ and the energy $E$ is conserved, and solving for ${\bf x}$, we now get
\be
\frac{d\tau}{dt}=\Delta = \frac{m}{E}\, \frac{1}{1-\frac{A_0 P_y}{E}}\ .
 \ee
 We can now compare the time rates for the particle at coordinates ${\bf x}_{1}$ and ${\bf x}_{2}$.
We obtain
\be
\frac{\delta t_2}{\delta t_1} \approx  1+\frac{2\kappa P_y}{E}\, {\bf E}\cdot ({\bf x}_1
-{\bf x}_2)+...\ 
\label{pooo}
\ee
The correction corresponds to the electromagnetic term in
formula \eqref{guare} (in $\eqref{guare}$, one has $|{\bf x}_1
-{\bf x}_2|=L$ and $|{\bf E}|=Q/r_0^2$). This is expected, since for $r_0\gg L$ the electromagnetic field is approximately constant over a distance $L$.
 
 $q\equiv 2\kappa P_y$ is the physical electric charge, since it appears as a coefficient in the electromagnetic coupling to KK matter. This is seen from the coupling $A_\mu \dot x^\mu$ in \eqref{ccc} and  is also seen in the covariant derivative in the field theory action \eqref{ftch}
 (upon the change $A_\mu\to 2\kappa A_\mu $ required to have the standard normalization in the kinetic term $-\frac14 F_{\mu\nu}F^{\mu\nu}$).

 Therefore,
 the formula \eqref{pooo} can be written as 
\be
\frac{\delta t_2}{\delta t_1} \approx  1+\frac{q}{E}\, {\bf E}\cdot ({\bf x}_1
-{\bf x}_2)+...\ 
\label{pooh}
\ee
where $E$ is the  energy of the particle.
This  dictates the change in the relative lifetimes of two identical  KK particles separated a distance $L=|{\bf x}_1
-{\bf x}_2|$ in a uniform electric field.
Note that the dependence on the KK radius is absorbed in $q$, but there is still dependence on $R$ through the energy
$E=\sqrt{m^2+n^2/R^2}$. We shall see later the significance of this fact. 

Thus far we have assumed a $4+1$ dimensional spacetime. It is interesting to see how the time rate formulas described above
generalize in more general Kaluza-Klein constructions.
In  attempts for a realistic Kaluza-Klein unification \cite{Horvath:1977st,Witten:1981me,Salam:1981xd,Mecklenburg:1983uk},
one assumes a $4+k$ dimensional space, where the ground state is $M_4\times Y_k$, where $M_4$ is the four-dimensional 
Minkowski space and $Y_k$ is a $k$-dimensional compact space. Non-abelian gauge symmetries then arise
from symmetries of $Y_k$. Let us assume that the coordinates of the internal space are $y^i,\ i=1,\cdots, k$,
and $T^ a$ the generators of the symmetry group $G$ of $Y_k$, which act on the internal coordinates
as $y^i\to y^i +K^i_a(y)$, where $ K^i_a(y) $ is the Killing vector corresponding to the symmetry generator $T^a$.
The $4+k$ dimensional metric now takes the form
\cite{Salam:1981xd}
\be
ds^2=g_{\mu\nu}(x) dx^\mu dx^\nu + h_{ij}(y) \big(dy^i- dx^\mu A_\mu^ a K_a^i(y)\big)\big(dy^j- dx^\nu A_\nu^ b K_b^j(y)\big)\ .
\label{nonabeg}
\ee
Assuming that the Standard Model gauge group is contained in $G$, {\it i.e.} 
$$
SU(3)\times SU(2)\times U(1)\subset G\ ,
$$
then a particular linear combination of generators in the Cartan subalgebra $\mathfrak{h}$ of $G$ generates the electromagnetic $U_{\rm e.m.}(1)$ symmetry.
With a convenient choice of basis, we may call $T^1$ the generator of $U_{\rm e.m.}(1)$. This
act on $y^1$ as a shift, $y^1\to y^1+{\rm const.}$
The electromagnetic charge is then associated to the conserved momentum $P_{y^1}$.

 Leptons and quarks of the Standard Model should be massless,
 that is, they should correspond to zero modes of the Dirac operator acting on the internal space.
One assumes that there is a massless spinor $\Psi$ in $4+k$ dimensions satisfying the Dirac equation \cite{Witten:1981me,Mecklenburg:1983uk,Mecklenburg:1981ju}
\be
\Gamma^A D_A \Psi =0\ ,\ \qquad A=1,\cdots, 4+k\ .
\ee
This decomposes 
\be
\slashed{D}^{(4)} \Psi + \slashed{D}^{\rm (int)} \Psi = 0\ ,\qquad \slashed{D}^{(4)}=\Gamma^\mu D_\mu\ ,\ \ \slashed{D}^{\rm (int)}=
\Gamma^i D_i\ . 
\ee
In four dimensions, the eigenvalues of $ \slashed{D}^{\rm (int)} $ are therefore seen as fermion mass.
Because $Y_k$ is a compact space, the spectrum of  $\slashed{D}^{\rm (int)} $ is discrete. The non-zero eigenvalues are of order $1/R$, where $R$ is the size of the compact space (typically Planckian size).
Thus, in order to obtain the Standard model spectrum, one has to look at zero modes of the Dirac operator.

Solutions of the zero mode equation form representations of the symmetry group $G$, since the operator $\slashed{D}^{\rm (int)}$ is $G$-invariant by construction.
The branching rules in the embedding
\be
 SU(3)\times U_{\rm e.m.}(1)\subset G
\ee
then determine the electromagnetic charges of the zero modes. It is important to note that, in this construction, the rest masses of the Standard Model particles originate from quantum
effects and therefore they are not proportional to the charge times $1/R$
as in the $4+1$ dimensional case. 
Zero-mode particles must carry the quantum numbers of Standard model particles and therefore they can have non-zero electromagnetic charges.

In higher dimensions  a fermion field will generally have non-trivial dependence on the internal dimensions, 
but will still  be a zero mode if it satisfies $\Gamma^i D_i \Psi=0$. This is due to a balance with the
scalar curvature of the internal space $R$, as seen from the  identity \cite{Mecklenburg:1983uk}
\be
(\Gamma^i D_i)^2 \Psi = \left(-D_iD^ i+\frac14 R\right)\Psi\ .
\ee
where $R$ is the curvature of the internal space.

Consider now a Standard Model fermion moving in geodesics in $4+k$ dimensions, in the presence
of an electric potential $A_0$. As in the 4+1 dimensional case, we may ignore gravitational back reaction
for any reasonable electric field produced in an Earth laboratory. As in previous examples, for generality we assume that the particle has a mass $m$ in $4+k$ dimensions
(taking then the $m=0$ limit is straightforward).
The worldline action for the particle is obtained from the metric \eqref{nonabeg}.
The calculation may  appear to be more complicated due to its dependence on the specific details of the internal space
$Y_k$, in particular,
on the metric $ h_{ij}$. 
However, in a suitable frame, the components of $h_{ij}$ are independent of the Cartan direction $y^1$
and the associated Killing vector is simply $\partial_{y^1}$. Then a configuration with $U_{\rm e.m.}(1)$ charge is generated by geodesic motion with nonzero $\dot y_1$.
This leads to essentially the same  time rate formulas as \eqref{pooo},
as  illustrated by the following  classical example.

Consider an internal seven-dimensional space
$$
Y_7 = S^5\times S^2\ .
$$
This has symmetry group $O(6)\times SU(2)$ and it includes
the Standard Model as a subgroup. 
We choose the metric
\bea
ds^2_7 &=&R_1^2 \left(d\theta^2+\cos^2\theta\ d\psi^2+\sin^2\theta \ d\varphi_1^2+
\cos^2\theta\sin^2\psi\ d\varphi_2^2+\cos^2\theta\cos^2\psi \ d\varphi_3^2\right)
\nonumber\\
&+&R_2^2\left(d\alpha^2+\sin^2\alpha\  d\varphi ^2\right)\ .
\eea
where $R_1$ and $R_2$ are the radii of $S^5$ and $S^2$ respectively.
A geodesic of a point-like particle can carry four independent Cartan charges,
three angular momenta in $S^5$ (associated with $\dot \varphi_{1},\ \dot \varphi_{2},\ \dot \varphi_{3}$)
and one angular momentum in $S^2$ (associated with $\dot \varphi$).
For the sake of computing the formula for the time rate
in the presence of an electric field, we
 may identify the $U_{\rm e.m.}(1)$ symmetry as the one generated by $\partial_{\varphi_1}$ and consider 
 geodesic motion in the equatorial plane $\theta=\frac{\pi}{2}$, $\psi=0$, with $\varphi_1=\varphi_1(t)$ and constant values for the other angles. Other choices lead to similar results.
Turning on the electromagnetic field $A_0$,  for this geodesic the nonabelian Kaluza-Klein ansatz \eqref{nonabeg}
gives the following  particle action 
\be
S= -m\int dt\ \Delta \ ,
\nonumber
\ee
\be
\Delta \equiv \sqrt{ 1- \dot {\bf x}^2 -\big(R_1 \dot \varphi_1 -2\kappa  A_0  \big)^2
}\ .
\label{accionmas}
\ee
The action is essentially the same as in the 4+1 dimensional case \eqref{acciones}. Following the same steps, we now get
the rate 
\be
\frac{\delta t_2}{\delta t_1} \approx  1+\frac{2\kappa J_1}{E}\, {\bf E}\cdot ({\bf x}_1
-{\bf x}_2)+...\ 
\label{poonn}
\ee
which is essentially the {\it same} time rate as \eqref{pooo}, now in terms of a conserved charge $J_1=n/R_1$ conjugate to $\varphi_1$.
The physical electric charge  is $q=2\kappa J_1=2n\kappa/R_1$. It is uniquely determined 
by the coupling to the electromagnetic field (in generic Kaluza-Klein
compactifications, the electric charge
is always proportional to $\kappa/R$; see
\cite{Witten:1981me,Salam:1981xd,Mecklenburg:1983uk}).

Considering now more general spaces, it is evident that the time rate of a Standard model particle will be affected by the presence of $A_0$, since it is part of the metric.
Generalizing the above calculation, in general one expects a linear dependence of the form 
\be
\frac{\delta t_2}{\delta t_1} \approx  1+\beta\ \frac{q}{E}\, {\bf E}\cdot ({\bf x}_1
-{\bf x}_2)+...\ 
\label{poohh}
\ee
where $\beta$ is a numerical constant and $q,\ E$ represent charge and the  energy of the particle, respectively.

It is worth noting  a crucial difference with respect to the $4+1$ dimensional case, where there are no fermion zero modes and $E=\sqrt{m^2+P_y^2}\geq n/R$. This inequality implies $\frac{q}{2E} \leq 2\kappa$.
Since $\kappa$ represents the Planck length, one thus needs a Planck scale
electric field in order to have a non-negligible effect in \eqref{pooh}.
However, for fermion zero modes the ratio ``charge/energy"  is not
bounded by $\kappa $.  This observation will be important
below.
	

High-energy accelerators involving muons offer a promising experimental setup  to test these effects.
The muon is unstable and can be used as a clock to measure the time dilation.
 The lifetime of a muon at rest is \cite{ParticleDataGroup:2022pth}
 \be
\tau_\mu= 2.1969811 \pm 0.0000022 \ {\rm s}\ .
 \ee
The time dilation predicted by special relativity has been measured
many times for atmospheric muons arising from cosmic rays
(for a review, see \cite{Gorringe:2015cma}).
High-energy accelerator experiments have also been carried out to verify time dilation in  straight paths \cite{Bailey:1977de}.
At CERN muon storage ring, a version of the twin paradox has been tested by comparing the lifetime of the circulating muon 
with the lifetime of the muon  at rest. The circulating muon
is subject to centripetal acceleration and has a longer lifetime than the muon at rest.

 A number of experiments are being planned to produce muon collisions at high luminosity and 10 TeV center-of-mass energy by the recently-formed International Muon Collider Collaboration
\cite{Accettura:2023ked}.  
An accurate determination of the  corrections to the muon lifetime is  crucial  for the planned muon colliders, as the number of  available muons for the experiment decreases exponentially with time. Having a sufficient amount of muons is one of  the central challenges faced by muon colliders.

The gravitational correction to time dilation is too small to be observed
with the current experimental accuracy.
Taking $r_0$ to be the Earth radius, one has $MG/R^2 \approx 9.8\, $m/s$^2$. Restoring the speed of light, the correction to the time rate due to gravitation is of order $(9.8\, {\rm m/s}^2)\, L/c^2\sim 10^{-15}(L/{\rm m})$. This is a very small effect. 
However, in laboratory experiments, the acceleration produced by electric fields is many orders of magnitude greater than the gravitational acceleration $MG/R^2$.
Consider the Fermilab muon $g-2$ experiment \cite{Muong-2:2023cdq}, where  muons of energy 3 GeV are injected in a ring 50 meters in circumference. A time rate of the form  \eqref{poohh} will give a correction to the relativistic time dilation of order 
$\frac{q}{E} |{\bf E}| L\sim 1$ already for an   electric field of $600$ kV/cm. But this effect has not been observed in \cite{Muong-2:2023cdq}, where the time-dilated lifetime of muons carries only the relativistic factor $\gamma$.
Thus, a correction to the time dilation of the form \eqref{poohh} inherited from Kaluza-Klein theories
would be in tension with  muon experiments. 
This is a powerful constraint for the construction of realistic Kaluza-Klein models. 
Although the analysis of fermions in realistic KK compactifications is beyond the scope of this note (see \cite{Salam:1981xd,Horvath:1977st} for discussions),  corrections of the form \eqref{poohh}  are expected on general grounds, since charged fermions must have interactions with the Kaluza-Klein electromagnetic field and this is a component of the 
higher-dimensional metric, therefore contributing to the time dilation formula \eqref{uno}.


\smallskip

In conclusion, this paper  explored  an alternative approach for probing extra dimensions. The method focuses on distinctive corrections that characterize time dilation for Kaluza-Klein particles. 
Although the present  model neglects relevant aspects that are needed in order to have a realistic Kaluza-Klein compactification, we believe that this area of research is promising and deserves thorough investigation.

\bigskip

\noindent\textbf{Acknowledgments}: This work is supported by 
 a MINECO grant PID2019-105614GB-C21.

%

\end{document}